\documentstyle[12pt,epsf,epsfig,wrapfig]{article}

\begin{document}

\begin{center}

{\bfseries SPIN EFFECTS AND ROLE OF CONSTITUENT QUARK SIZE}

\vskip 5mm

\underline{S.M. Troshin},
N.E. Tyurin

\vskip 5mm

{\small
 {\it

Institute for High Energy Physics,\\
 Protvino, Moscow Region, 142281, Russia

}}

\end{center}

\vskip 5mm

\begin{abstract}

The nonperturbative mechanism for spin effects in inclusive production
based on the constituent
quark model is considered. The main role belongs to the orbital angular momentum and
 the polarization of the $\bar q q$--pairs in the internal structure
  of the constituent quarks.

\end{abstract}

\vskip 8mm
It is well known fact that spin observables demonstrate rather complicated behaviour
and  it somehow is related to a scarcity of the experimental data in the field.
However, it seems very important to find some general trends in the data
even under such strained circumstances. One among the more or less stable
 patterns is the momentum
transfer dependence of various spin observables in inclusive production. To be
specific
we will concentrate on a particular problem of $\Lambda$--polarization. Experimentally,
the situation is stable and clear. $\Lambda$--polarization is negative and energy
independent. It grows linearly with $x_F$ for $p_\perp > 0.8$ GeV/c, and for large
values of the momentum transfer ($0.8<p_\perp < 3.5$ GeV/c) it is
$p_\perp$--independent \cite{heller,newrev}. It is remarkable that both parameters $A_N$ and $D_{NN}$ show
$p_\perp$--dependence similar to polarization \cite{bravar}.

 In perturbative QCD.
a straightforward collinear factorization leads to very small values of $P_\Lambda$
\cite{pump,gold}.
pQCD modifications and in particular account for higher twists result in the dependence
$P_\Lambda\sim 1/p_\perp$ \cite{efrem,sterm,koike}. This behavior still does not
correspond to the experimental
trends. Account for $k_\perp$--effects when the source of polarization is moved into the
polarizing fragmentation functions also leads to falling $P_\Lambda\sim k_\perp/p_\perp$
at large $p_\perp$ values \cite{anselm}. Potentially $\Lambda$--polarization could
 become even a more serious problem
 that the nucleon spin problem. And in any case the both problems are interrelated.

The essential point here is that the vacuum
at short distances is taken to be a
 perturbative one.
However,  polarization dynamics could have its roots hidden in the genuine
nonperturbative sector of QCD. The models exploiting confinement and the chiral symmetry
breaking have been proposed. Our model considerations \cite{pollam}
are based on the effective Lagrangian
approach which in addition to the four--fermion interactions of the original NJL model
 includes the six--fermion $U(1)_A$--breaking term.

 Chiral symmetry breaking generates quark masses:
 \[
  m_U = m_u-2g_4\langle 0|\bar u u|0\rangle-2g_6\langle 0|\bar
d d|0\rangle \langle 0|\bar s s|0\rangle.
\]
In this approach massive quarks appear as quasiparticles,
 i.e. current quarks surrounded by a cloud
of quark--antiquark pairs of different flavors. For example, for the $U$--quark the ratio
\[
\langle U|\bar s s|U\rangle/\langle U|\bar u u+\bar d d+\bar s s|U\rangle
\]
is estimated as $0.1-0.5$. The scale of spontaneous chiral symmetry breaking is
\[
\Lambda\simeq 4\pi f_{\pi}\simeq 1\quad\mbox{GeV/c}
\]
and provides the momentum cutoff which determines a transition to the partonic picture.
We consider nonperturbative hadron as consisting of constituent quarks located
in the central part of the hadron and embedded into a quark condensate.

Respectively, spin of the constituent quark is given by the following "spin balance
equation":
\begin{equation}\label{bal}
 J_{U}=1/2  =  S_{u_v}+S_{\{\bar q q\}}+ L_{\{\bar qq\}}=
               1/2+S_{\{\bar q q\}}+ L_{\{\bar qq\}},
\end{equation}
where $L_{\{\bar qq\}}$ is the orbital angular momentum of quark--antiquark
pairs in the structure of the constituent quark. Its value
 can be estimated \cite{pollam}
with the use of the polarized DIS data:
\begin{equation}\label{lv}
L_{\{\bar qq\}}\simeq 0.4
\end{equation}
This means that the cloud quarks rotate coherently and significant part of the
constituent quark spin is to be associated with the orbital angular momentum.
In the model just this orbital motion of quark matter is the origin of asymmetries
in inclusive processes. It is to be noted that the only effective degrees of freedom
here are quasiparticles. The gluon degrees of freedom are overintegrated,
and the six-fermion operator in the NJL Lagrangian  simulates the effect
of the gluon operator $({\alpha_s}/{2\pi})G^a_{\mu\nu}\tilde G^{\mu\nu}_a$ in QCD.
It is also important to note the exact compensation between the spins of
$\bar q q$-pairs and their orbital momenta:
\begin{equation}\label{comp}
L_{\{\bar qq\}}=-S_{\{\bar qq\}},
\end{equation}
which follows from Eq. (\ref{bal}).

Assumed picture of hadrons implies that overlapping and interaction of peripheral
clouds and condensate excitation occur at the first stage of the collision.
 As a result
massive virtual quarks appear in the overlapping
 region and some mean field is generated.
Inclusive production of hyperon results from the two mechanisms:
recombination of constituent quark with virtual massive strange quark
(soft interactions) or from the constituent quark scattering in the mean
field, its excitation and appearance of a strange quark as a result of decay of
the parent constituent quark. The second mechanism is determined by
 interactions at the distances smaller than the constituent quark
 radius ($r<R_Q\sim 1/\Lambda_\chi$) and is associated with hard interactions.
Thus, we adopt a two-component picture of hadron (hyperon) production which
incorporates interactions at long and short distances and it is the short
distance dynamics which leads to production of polarized $\Lambda$'s.

Polarization of a strange quark results from the multiple scattering of parent
constituent quark $Q$ in the mean field where it gets polarized
\begin{equation}
 {\cal{P}}_Q\propto -I\frac{m_Qg^2}{\sqrt{s}}\label{xpl}
\end{equation}
and the polarization is nearly constant in the model since $m_Q\sim m_h/3$ and
$I\sim \sqrt{s}$. The second crucial point is correlation between $s$--quark
 polarization and polarization of the parent quark $Q$. Indeed, the total orbital
 momentum of $\bar q q$--pairs in the constituent quark which has polarization
${\cal{P}}_Q(x)$ is
\begin{equation}\label{pq}
L_{\{\bar qq\}}^{{\cal{P}}_Q(x)}={\cal{P}}_Q(x)L_{\{\bar qq\}},
\end{equation}
where the value $L_{\{\bar qq\}}$ on the right hand side enters Eq. (\ref{bal})
written for the constituent quark with polarization $+1$. On the basis
 of Eq. (\ref{comp}) we suppose that there is a compensation between spin and
 orbital momentum of strange quarks inside the constituent quark

\begin{equation}\label{comps}
  L_{s/Q}= -J_{s/Q}=\alpha {\cal{P}}_Q(x)L_{\{\bar qq\}},
\end{equation}
where the parameter $\alpha$ determines the fraction of orbital momentum
due to the strange quarks.
Eq. (\ref{comps}) is quite similar to the conclusion made in the framework
of the Lund model but has a different dynamical origin rooted in the mechanism
of the spontaneous chiral symmetry breaking.

Final expression for the polarization is
\begin{equation}\label{pol}
P(s,x,p_{\perp})=\sin [\alpha {\cal{P}}_Q(x)L_{\{\bar qq\}}]
\frac{R(s,x,p_{\perp})}{1+R(s,x,p_{\perp})}.
\end{equation}
The function $R$ is the cross--section ratio of hard and soft processes.
At $p_\perp >\Lambda_\chi$ the function $R(s,x,p_{\perp})\gg 1$ and the polarization
saturates
\begin{equation}\label{pol1}
P(s,x,p_{\perp})=\sin [\alpha {\cal{P}}_Q(x)L_{\{\bar qq\}}].
\end{equation}
{\it Characteristic $p_\perp$--dependence of $\Lambda$--polarization follows from
Eqs. (\ref{pol}) and (\ref{pol1}): polarization is vanishing for $p_\perp <\Lambda_\chi$,
it gets an increase in the region of $p_\perp\simeq \Lambda_\chi$ and polarization saturates
and becomes $p_\perp$--independent (flat) for $p_\perp >\Lambda_\chi$. The respective
scale is the scale of the spontaneous chiral symmetry breaking $\Lambda_\chi\simeq 1$ GeV/c.}
Such a behavior of polarization follows from the fact that constituent quarks themselves
have slow (if at all) orbital motion and are in the $S$--state, but interactions with
$p_\perp >\Lambda_\chi$ resolve the internal structure of constituent quark and feel
the presence of internal orbital momenta inside this constituent quark.

To describe the data quantitatively we have to introduce some parameterization.
 The function
\begin{equation} R(s,x,p_\perp)=
C(x){\exp(p_\perp/m)}/{(p_\perp^2+\Lambda_\chi^2)^2}  \label{rat}
\end{equation}
implies typical behavior of cross--sections for hard and soft processes.
The form
\[
{\cal{P}}_Q(x)={\cal{P}}_Q^{max}x
\]
is suggested by the ALEPH data \cite{aleph}. The value of Eq. (\ref{lv})
alongside with $m=0.2$ GeV and
$\alpha=0.8$ provides a rather good fit to the experimental data (Fig. 1 and 2).
\begin{figure}[h]
\begin{center}
\includegraphics[height=.2\textheight]{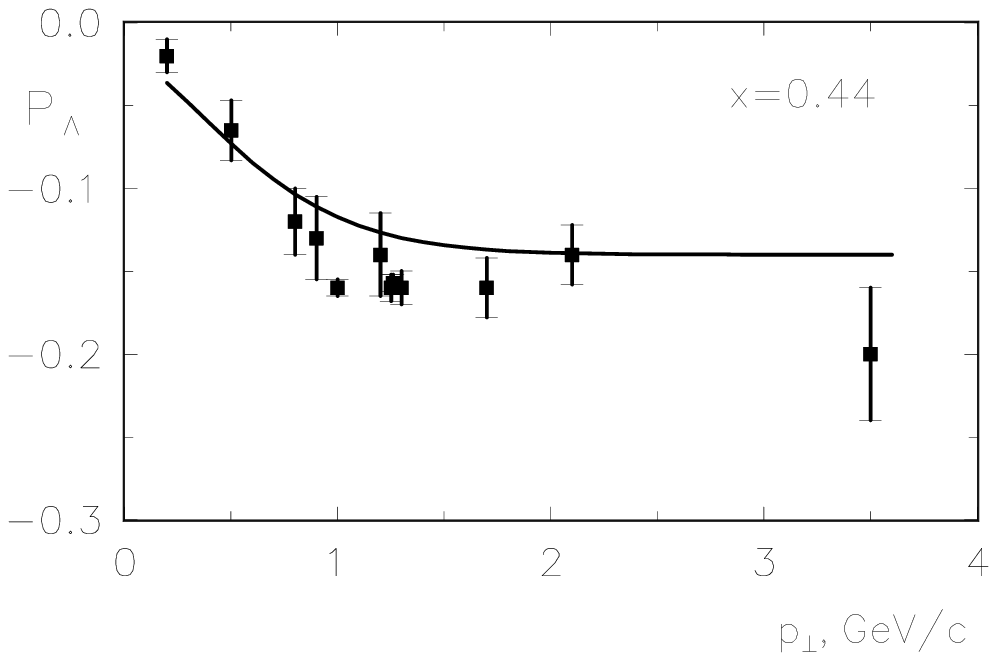}
\includegraphics[height=.3\textheight]{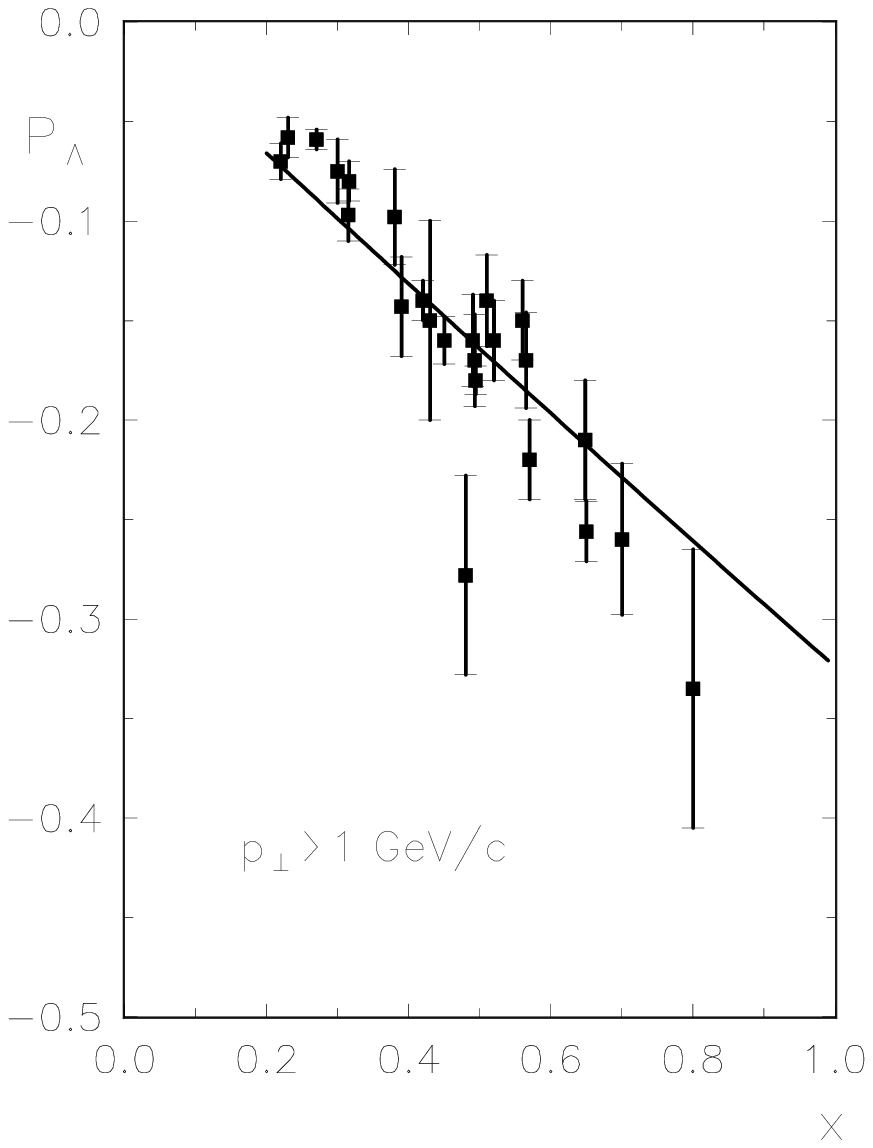}
\end{center}
\caption{Transverse momentum (left) and Feynman $x$ (right) dependencies of $P_\Lambda$.}
\label{fig:1}
\end{figure}

{\it In the model the spin transfer parameter $D_{NN}$ is positive since
$P_\Lambda$  has the same sign as ${\cal{P}}_Q$. The model
also predicts similarity of $p_\perp$ dependencies for the different spin
observables. The respective features were clearly seen
 in E-704 experiment \cite{bravar}.}
\begin{center}
{\bf Experimental prospects}
\end{center}
In this approach the short distance interaction with $p_\perp\> \Lambda_\chi$
observes a coherent rotation of correlated $\bar q q$--pairs inside the
 constituent quark and not a gas of the free partons. The nonzero internal orbital
 momenta in the constituent quark means that there are significant multiparton
 correlations. The important point is what the origin  of this orbital angular
 momentum is. The analogy with an anisotropic extension of the theory
 of superconductivity seems match well with the adopted picture for a constituent
 quark. An axis of anisotropy can be associated with the polarization vector
 of the valence quark located at the origin of the constituent quark.

 It seems interesting to perform $\Lambda$--polarization measurements at RHIC. When two
 polarized nucleons are available one could measure three--spin correlation parameters
 $(n,n,n,0)$ and $(l,l,l,0)$ in the processes
\begin{equation}
p_{\uparrow,\rightarrow}+p_{\uparrow,\rightarrow}=\Lambda _{\uparrow,\rightarrow}+X.
\end{equation}
It would provide important data to study  mechanisms of hyperon polarization.

Experimentally observed persistence and constancy of $\Lambda$--hyperon
polarization
 means
 that chiral symmetry is not restored in the region
 of energy and values of $p_{\perp}$ where experimental measurements
 were performed. Otherwise we would not
 have any constituent quarks and should expect
 a vanishing polarization of $\Lambda$. It is interesting
 to perform  {\it $\Lambda$--polarization measurements at RHIC and the LHC.
 It would allow to make a direct check of perturbative QCD and allow to
 make a cross-check of the QCD background estimations based on
 perturbative calculations for the LHC}.
On the basis of the above model we expect significant $P_\Lambda$ at RHIC energies.
On the base of the model in   one  expects
 zero polarization in the
region where QGP  has  formed, since  chiral symmetry is restored
and there is no room for  quasiparticles such as  constituent quarks.
The absence or strong diminishing of transverse  hyperon polarization
can be used
therefore as a signal of QGP formation in heavy-ion collisions.
This prediction should also be valid
for the  models based on confinement, e.g. the Lund and Thomas precession
model. In particular, the polarization of $\Lambda$ in heavy--ion collisions
in the model based on the Thomas precession was described in
where nuclear effects were discussed as well.
However, we do not expect a strong diminishing of the $\Lambda$--polarization due to
the nuclear effects:  the available data show a weak $A$--dependence and are
not sensitive to the type of the target. Thus, we could use a vanishing polarization
of $\Lambda$--hyperons in heavy ion collisions as a sole result of QGP formation provided
the corresponding observable is non-zero  in  proton--proton collisions.
 The prediction based on this observation
would be a decreasing behavior of polarization of $\Lambda$ with the impact parameter
in heavy-ion collisions in the region of energies and densities where QGP was
produced:
$P_\Lambda(b)\to 0\quad \mbox{at}\quad b\to 0$,
since the overlap is maximal at $b=0$. The value of the impact parameter  can be
controlled by the centrality in heavy--ion collisions.
The experimental program should therefore
include  measurements of $\Lambda$--polarization
in $pp$--interactions first, and then if a significant polarization would be
measured,  the corresponding measurements could be a useful tool for the
 QGP detection.
Such measurements seem to be experimentally feasible
at  RHIC and  LHC provided it is supplemented with forward
detectors.

{ \small\it One of the authors (S.T.) would like to thank Organizing Committee of the
10-th Workshop on High Energy Spin Physics
(Nato Advanced Research Workshop) Spin-03, Dubna, September 16-20, 2003
for the valuable support and warm hospitality in Dubna.}


{\small
\begin{thebibliography}{99}
\bibitem{heller} G. Bunce at al.,
Phys. Rev. Lett. \bf 36 \rm,   1113 (1976);\\
For a history of the hyperon polarization discovery, see: T. Devlin, in SPIN 94,
 Proceedings of the
 11th International Symposium on High-energy Spin Physics
and the 8th International Symposium on Polarization Phenomena
 in Nuclear Physics, Bloomington, Indiana, 15-22 September 1994,
AIP Conf. Proc. Woobury, New York, 1995,  eds.  K. Heller and S. Smith, p. 354.
\bibitem{newrev}
L. Pondrom, Phys. Rep. \bf 122\rm ,  57 (1985);\\
K. Heller, in
 Proceedings of the
7th International Symposium on High-Energy Spin Physics, Protvino, Russia,
1987,  p. 81;\\
 J. Duryea et
al.,
Phys. Rev. Lett. \bf 67 \rm,  1193 (1991).\\ A. Morelos et al.,
Phys. Rev. Lett. \bf 71 \rm,   2172 (1993);\\
K. A. Johns et al., in SPIN 94, Proceedings of the
 11th International Symposium on High-energy Spin Physics
and the 8th International Symposium on Polarization Phenomena
 in Nuclear Physics, Bloomington, Indiana, 15-22 September 1994,
AIP Conf. Proc. Woobury, New York, 1995,  eds.  K. Heller and S. Smith,
 p. 417;\\
L. Pondrom, ibid., p. 365;\\
A. D. Panagiotou, Int. J. Mod. Phys A \bf 5\rm, 1197 (1990).
\bibitem{bravar}
 A. Bravar, in SPIN 98 Proceedings of the
13th International Symposium on High-Energy Spin Physics, Protvino, Russia,
 8-12 September 1998, N. E. Tyurin, V. L. Solovianov, S. M. Troshin,
 and A. G. Ufimtsev, eds. (World Scientific, Singapore, 1999) p. 167.
\bibitem{pump}
G. L. Kane, J. Pumplin, and W. Repko,
Phys. Rev. Lett. \bf 41\rm, 1689 (1978).
\bibitem{gold}
W. G. D. Dharmaratna and G. R. Goldstein,
Phys. Rev. D \bf 53\rm, 1073 (1996)
\bibitem{efrem}
A. V. Efremov and O. V. Teryaev, Sov. J. Nucl. Phys. \bf 36\rm, 140 (1982).
\bibitem{sterm}
J. Qiu and G. Sterman, Phys. Rev. D\bf 59\rm, 014004 (1999).
\bibitem{koike}
Y. Kanazawa and Y. Koike, Phys. Rev. D,\bf 64\rm, 034019 (2001).
\bibitem{anselm}
M. Anselmino, D. Boer, U. D'Alesio, and F. Murgia, Phys. Rev. D.\bf 63\rm, 054029 (2001).
\bibitem{pollam}
S. M. Troshin and N. E. Tyurin, Phys. Rev. D\bf55\rm, 1265 (1997).
\bibitem{aleph} D. Buskulic et al. (ALEPH Collaboration), Phys. Lett. B\bf 374\rm, 319 (1996).
\end{thebibliography}}
\end{document}